\documentclass[twoside,usletter,11pt,leqno,onecolumn]{article}  
\usepackage{ltexpprt} 
\usepackage{enumerate}
\usepackage{rotating}
\usepackage{wrapfig}
\usepackage{amsmath}
\usepackage{graphicx}
\usepackage{amsmath}

\newcommand{\CA}[0]{$\mathrm{C}_\alpha$}
\newcommand{\calP}[0]{$\mathcal{P}$}
\newcommand{\calQ}[0]{$\mathcal{Q}$}

\newcommand{\calD}[0]{$\mathcal{D}$}
\newcommand{\calC}[0]{$\mathcal{C}$}

\newcommand{\myfig}[1]{Fig. \ref{#1}}
\newcommand{\mysection}[1]{Section \ref{#1}}
\newcommand{\mysectionpair}[2]{Sections \ref{#1} and \ref{#2}}
\newcommand{\myeqn}[1]{Equation \ref{#1}}

\newcommand{\myeqnspair}[2]{Equations \ref{#1} and \ref{#2}}

\newcommand{\mytab}[1]{Table \ref{#1}}

\newcommand{\ignore}[1]{}

\begin{document}
%
\title{Statistical Inference of a canonical dictionary of protein substructural fragments}
\setcounter{footnote}{1}
\author{Arun S. Konagurthu
\footnote{Clayton School of Computer Science and Information Technology, Monash University, VIC 3800 Australia.}~$^*$\\
\and 
Arthur M. Lesk, 
\setcounter{footnote}{2}
\footnote{The Huck Institute of Genomics, Proteomics and Bioinformatics and the
Department of Biochemistry and Molecular Biology, Pennsylvania State University,
University Park PA 16802 USA}
\and 
David Abramson,
$^\dagger$
\footnote{Research Computing Center, University of
Queensland, St Lucia, QLD 4072 Australia}
\and
Peter J. Stuckey,
\footnote{Department of Computing and
Information Systems, University of Melbourne, Parkville VIC 3010 Australia}
\and
Lloyd Allison~$^\dagger$\setcounter{footnote}{0}\footnote{Correspondence can be addressed to: arun.konagurthu@monash.edu or lloyd.allison@monash.edu}
}
\date{}

\maketitle

\begin{abstract} Proteins are biomolecules of life. They fold into a great
variety of three-dimensional (3D) shapes. Underlying these folding patterns are
many recurrent structural fragments or building blocks (analogous to `LEGO{\textsuperscript{\textregistered}}
bricks'). This paper reports an innovative statistical inference approach
to discover a comprehensive dictionary of protein structural building blocks from a large corpus of experimentally determined protein structures. Our approach is built on the Bayesian and information-theoretic criterion of minimum message length. To the
best of our knowledge, this work is the first systematic and rigorous treatment
of a very important data mining problem that arises in the cross-disciplinary area of structural bioinformatics. The quality of the dictionary we find is demonstrated by its explanatory power -- any protein within the corpus of known 3D structures can be dissected into
successive regions assigned to fragments from this dictionary. This induces a
novel one-dimensional representation of three-dimensional protein folding patterns, suitable for application of the rich repertoire
of character-string processing algorithms, for rapid identification of folding patterns of newly-determined structures.
This paper presents the details of the methodology used to infer the dictionary of building blocks, and is supported by illustrative examples to demonstrate its effectiveness and utility.  
\end{abstract}

\section{Introduction} 
\label{sec:intro}
Proteins are molecules central to life. They are
responsible for biological and cellular functions in all organisms. Each
protein folds into a three-dimensional (3D) shape, determined by the intrinsic
properties of its sequence or chain of amino acid residues. 
Among the major triumphs of modern science are the techniques to experimentally determine the 3D structures
of proteins at atomic resolution. Worldwide structure determination efforts
have resulted in a fast growing public database, the protein
data bank (wwPDB)~\cite{wwpdb}. Currently wwPDB contains atomic coordinates of $91,550$ experimentally solved protein structures, whose size is doubling every five years. This database provides a rich source of structural and architechtural information for knowledge discovery and data mining applications that contribute to the advances made in life sciences in medicine.

Understanding the architectural principles of protein 3D structure is
fundamental to biological research. It is  well known that protein folding
patterns contain recurrent structural themes, commonly helices and pleated
sheets~\cite{pauling1951structure,pauling1951pleated}. However the identification of a canonical set of building blocks
(analogous to LEGO{\textsuperscript{\textregistered}} bricks) of protein structures still remains an important
open questions in biology.

Previous investigations have sought to identify a dictionary of fragments as building blocks of protein structures~\cite{unger89,wodak1990,comproux1999,micheletti2000,kolodny02,godzik05,joseph10}. (By fragment we mean a contiguous region within the folding pattern of a protein -- this can be viewed as a 3D analogue of a 1D substring.) However, these approaches largely rely on \textit{ad hoc}
clustering of short fragments of some fixed-length, usually 4 to 10
amino acid residues long. The restriction of generating fixed-length 
fragment libraries is an artificial constraint, 
mainly employed to work around the difficulty of the search problem that manifests when trying to identify recurring fragments of arbitrary length. Thus, the question, \textit{what is the canonical dictionary of fragments (of arbitrary lengths) of which all proteins are made}, remains fundamentally unsolved.

This paper addresses the above question by framing it as as a statistical inference problem. Our approach relies on the Bayesian method of minimum message length inference~\cite{wallace1969,wallace2005}, where the optimal fragment dictionary is defined objectively as the one which permits the most concise explanation
(or technically, shortest lossless encoding) of the coordinates of a collection
of source protein structures.
To the best of our knowledge, our work is the first objective and systematic treatment of this important question, addressed using a statistically rigorous  approach which investigates the compressibility of protein coordinate data.

We mine these building blocks from a collection of 8992 experimentally determined protein structures, whose coordinates were solved at atomic resolution, and available from the wwPDB~\cite{wwpdb}. These source structures are dissimilar in amino acid sequence to avoid experimental and selection bias observed within the wwPDB. In other words, the collection we use is comprehensive and unbiased, representative of all known protein folding patterns in the wwPDB.

Our approach discovered 1711 fragments or building blocks, ranging in length
from 4 to 31 amino acid residues. This dictionary allows the efficient,
lossless representation (or encoding) of any given protein structure.  For a
particular protein structure, the optimal lossless encoding contains a
dissection (or segmentation) -- that is, a designation of successive
non-overlapping regions in the protein structures that match the assigned
dictionary fragments -- and a statement of spatial deviations (or corrections)
that should be applied to the coordinates of each assigned dictionary fragment
so that the coordinates corresponding to each region in the actual structure
can be recovered losslessly. We note that the regions that do not efficiently
match any dictionary fragment are assigned to a `null model'; in these cases,
the spatial deviations bear the entire weight of the description -- this is
tantamount to stating the coordinates of the region in the source structure 
raw (or \textit{as is}).

The organization of the paper is as follows. \mysection{sec:mml} gives a brief
introduction to the minimum message length criterion. \mysection{sec:dictmml}
provides the foundations of the dictionary inference problem using the MML
framework. This  involves designing transparent
communication processes, developing lossless encoding strategies, and evolving efficient search algorithms. \mysectionpair{sec:nullmodel}{sec:desc_protein} give these details. \mysectionpair{sec:describecollection}{sec:searchdictionary} 
provide further technical details to compute the total explanation message 
length, and the strategy that allows us to  identify the 
best dictionary of fragments.  The paper concludes  with \mysection{sec:results} providing various results including illustrative examples of the effectiveness and the utility of the dictionary we discover.

\section{Minimum Message Length Framework}\label{sec:mml} Minimum Message
Length (MML) \cite{wallace1969,wallace2005} is a hypothesis (or model)
selection paradigm which links statistical inference with information theory
and data compression. 

MML is a Bayesian method of inference.  Formally, let $E$ denote a mass of
observed data (or evidence) and $H$ a hypothesis on the data. From Bayes's
theorem~\cite{BayesTheorem} we have: $P(H\&E) = P(H)\times P(E|H)= P(E)\times
P(H|E)$, where 
$P(H)$ is the \textit{prior} probability of hypothesis $H$,
$P(E)$ is the prior probability of data $E$,
$P(H|E)$ is the \textit{posterior} probability of $H$ given $E$, and 
$P(E|H)$ is the \textit{likelihood}. 

In the Bayesian framework, two competing hypotheses can be compared using the
ratios of their posterior probabilities:
$$
\frac{P(H_1|E)}{P(H_2|E)} = \frac{P(H_1)P(E|H_1)}{P(H_2)P(E|H_2)}
$$
Usually, the goal of inference is to select the hypothesis with the highest
posterior probability.

MML offers a complementary view of Bayesian inference by linking the
probability of an event with the message length required to transmit (or
communicate, explain, describe) it \textit{losslessly}.  The mathematical
theory of communication~\cite{Shannon48} gives the relationship between the
message length $I(e)$ to communicate an event $e$ losslessly, and its
probability $P(e)$: $I(e) = -\log P(e)$.\footnote{The unit of measurement of
information depends on the base of the logarithm; $\log_2$ gives message
lengths measured in bits, while $\ln$ gives the same measured in nits.}

Therefore, by applying Shannon's insight to Bayese theorem above, we get:
$$I(H\&E) = I(H) + I(E|H) = I(E) + I(H|E)$$ Similarly, two competing hypotheses
can be compared as:
$$
I(H_1|E)-I(H_2|E) = I(H_1)+I(E|H_1) - I(H_2)-I(E|H_2)
$$
It directly follows that the  best hypothesis $H^*$ is the one for which the
expression  $I(H^*)+I(E|H^*)$ takes the \textit{minimum} value.  (Notice, this
is equivalent to maximizing the posterior probability of the hypothesis given
the data, $P(H^*|E)$.)

MML is best understood as a communication process between an
imaginary pair of transmitter (Alice)  and receiver (Bob) connected by a
Shannon channel.  Alice's objective is to send the data $E$ using an
explanation message in a form that Bob can receive it losslessly.  Alice and
Bob agree on a \textit{ codebook} containing  general rules of communication
composed solely of common knowledge about typical, hypothetical data.  Anything
that is not a part of the codebook must be strictly transmitted as a part of
the explanation message.  If Alice can find the best hypothesis, $H^*$ on the
data, Bob will receive a decodable explanation message most economically.

Alice sends the explanation message of $E$ in two parts. In the first part she
transmits the best hypothesis, $H^*$, she could find on the data $E$ taking
$I(H^*)$ bits to communicate. In the second, Alice transmits the details of the
observed data $E$ not explained by the hypothesis $H^*$, taking $I(E|H^*)$ bits
to communicate.  (That is, this part correspond to the deviations of the
observed data $E$ with respect to $H^*$).  Notice that MML inference gives a
natural trade-off between hypothesis complexity ($I(H^*)$) and quality of its
fit to the data ($I(E|H^*)$).

\section{Inferring the dictionary using the MML criterion}
\label{sec:dictmml}
\noindent\textit{Preliminaries:} Let $\mathcal{C}$ denote a \textit{collection}
of source protein structures $\{\mathcal{P}_1, \mathcal{P}_2, \cdots,
\mathcal{P}_{|\mathcal{C}|}\}$. In this work we use a subset of 8992 structures
from the protein data bank  after removing amino acid sequence bias. That is,
no two structures in the collection \calC\ have a sequence similarity greater
than $40\%$.

Any protein structure $\mathcal{P}$ is represented as an ordered list of
$(x,y,z)$ coordinates of its alpha Carbon (\CA) atoms along the protein
backbone, denoted here as $\mathcal{P}=\{p_1,\cdots,p_{_{\scriptsize
|\mathcal{P}|}}\}$. All protein coordinates are defined in Angstrom units ($1
$\AA$=10^{-10}$ meters.)

Let $\mathcal{D} =\{\mathcal{Q}_1, \mathcal{Q}_2, \cdots,
\mathcal{Q}_{|\mathcal{D}|}\}$ denote a \textit{dictionary} of fragments. Each
dictionary \textit{element} $\mathcal{Q}=\{q_1,\cdots,q_{_{\scriptsize
|\mathcal{Q}|}}\}\in\mathcal{D}$ is a substructural fragment (\textit{i.e.,} a
list of coordinates corresponding to a consecutive region) derived from some
$\mathcal{P}\in\mathcal{C}$, of arbitrary length
($|\mathcal{Q}|<|\mathcal{P}|$).  

We note that each $(x,y,z)$ comes specified (in the protein data bank) to 3
positions after the decimal place.  Since we are dealing with inference based
on lossless compression, we denote $\epsilon = 0.001$ as a parameter that
specifies the accuracy to which coordinate data should be stated.

\noindent\textit{Rationalizing this problem in the MML framework:} In this
work, any dictionary  \calD\ of fragments is a hypothesis of building blocks on
a collection of structures \calC, with its observed coordinates acting as
evidence for inference. Therefore, using the information-theoretic restatement
of Bayes's theorem describe in \mysection{sec:mml}, we get:
\begin{equation}
I(\mathcal{D}\&\mathcal{C}) = I(\mathcal{D}) + I(\mathcal{C}|\mathcal{D}).
\label{eqn:mainobjective}
\end{equation}

Rationalizing \myeqn{eqn:mainobjective} as a communication process between
Alice and Bob, the measure of quality of any proposed dictionary of
substructures is the total length of the explanation message that Alice
transmits to Bob so that all the coordinates in the collection of source
structures are received losslessly. Given that even unrelated proteins contain
common, recurrent fragments (or building blocks), Alice and Bob could
reasonably hypothesize that they could apply this observation to transmit the
coordinates of various structures more concisely, by using the dictionary of
building blocks as the basis of communication.  It is intuitive to see that the
better a dictionary of fragments in terms of how well they describe (i.e.,
\textit{fit}) the observed coordinates, the more economical is the description
of the source structures in the collection. 

To be useful for illuminating common building blocks over all proteins, the
dictionary must be the same for all structures; that is, the dictionary does
not change regardless of an individual structure that is being transmitted.
Before transmitting the coordinates of the collection, Alice first encodes and
sends Bob the canonical dictionary of fragments (taking $I(\mathcal{D})$ bits).
With this information, Bob has a dictionary of substructures but not the
coordinates of the source structures in the collection. Note again that Alice
needs to send the dictionary only once (as a header (or first part) of the
total explanation message): she need not restate it as part of the subsequent
encodings of coordinates of particular protein structures being transmitted.
Each subsequent message consists of the segmentation, that is, the optimal
assignment of successive regions in a protein structure to dictionary
fragments, plus the corrections (or vector deviations) required because each
region in the source protein deviates from its assigned dictionary element. This takes $I(\mathcal{C}|\mathcal{D})$ bits. 

In proposing a dictionary Alice and Bob face a tradeoff. They could use a
large, all-encompassing fragment dictionary whose elements fit regions of
proteins very well, leaving only small deviations in the assigned regions of
the proteins to be described. In this case the explanation message length for
each protein would be dominated by the explanation cost of the dictionary and
the assignment of dictionary fragments to regions, because a large dictionary
requires a larger message to explain itself and to nominate a dictionary
element. Alternatively, they could choose a small dictionary, in which case the
message length would be dominated by the transmission of the corrections
(vector deviations). As described in \mysection{sec:mml}, the MML criterion
provides an objective tradeoff between the dictionary complexity and its fit
with the coordinate data observed in the collection. 

\noindent\textit{Optimality criterion:} The optimal dictionary involves finding
a dictionary of fragments that minimizes the total message length equation
shown in \myeqn{eqn:mainobjective}. To achieve this involves the following
criteria:
\begin{enumerate}
\item Assume some dictionary \calD\ is given (containing arbitrary number of
fragments, each of arbitrary length). According to the MML framework, the
\textit{optimal encoding of a particular protein structure} \calP\ using the
specified dictionary \calD\  is defined as the combination, of minimal encoding
length, of assignments of successive non-overlapping regions in \calP\ to
fragments in \calD, plus statements of spatial deviations relative to each
assigned fragment per region to recover the observed coordinates in \calP\
losslessly.  Sections \ref{sec:nullmodel} and \ref{sec:desc_protein} present
all the technical details to achieve this.

\item Next, given the above method to optimally encode a particular protein
using a specified dictionary, the  \textit{optimal encoding of a collection of
protein structures} \calC, all using the same fixed dictionary \calD, requires
the one-off statement of the dictionary \calD\ (as a header to the subsequent
explanation message), plus the optimal encodings of each individual protein
$\mathcal{P}_i\in\mathcal{C}$ using the method in Step 1.  Section
\ref{sec:describecollection} presents the details to achieve the optimal
encoding of a collection with any specified dictionary.

\item Finally, given the method to optimal encode a collection of proteins in
Step 2, an \textit{optimal dictionary for the collection of protein structures}
can be objectively defined as the one for which the  one-off specification cost
of the dictionary \calD, plus the sum of the optimal encodings of all the
proteins $\mathcal{P}_i\in\mathcal{C}$, yields the \textit{shortest}
explanation message.  Section \ref{sec:searchdictionary} describes the search
to infer the optimal dictionary. 
\end{enumerate}

\section{Null model message length}
\label{sec:nullmodel}
The null model implies that that the coordinates are being transmitted raw.
However this message must not be wilfully inefficient. Recently, Konagurthu
\textit{et al.}~\cite{sst} introduced statistical models to encode protein
coordinate data, while dealing with a completely different problem of assigning
secondary structure to protein coordinates. This work builds on their encoding
schemes.

The null model encoding relies on the empirical observation that the distance
between successive \CA\ atoms in any protein chain is highly constrained at
around $3.8$ \AA\, with a small deviation about this mean. For a chain (or
ordered list) of coordinates $\{p_1,p_2,\dotsc,p_n\}$, any $p_i$ can be
transmitted raw given its previous coordinate $p_{i-1}$ as follows: First the
distance $r_i$ between $p_{i-1}$ and $p_i$ is encoded and transmitted over a
normal distribution $\mathcal{N}(r;\mu,\sigma)$ stated to $\epsilon$ accuracy,
with $\mu=3.8$ \AA\ and $\sigma=\pm0.2$ \AA. This takes 
\begin{equation}
\label{eqn:codelen_normal}
I_{\mathcal{N}}(r_i) = \log_2\left(\dfrac{\sqrt(2\pi)\sigma}{\epsilon}\right)
 + {\dfrac{(r_i-\mu)^2}{2\sigma^2}}\log_2 e ~~\mathrm{bits.}
\end{equation} 
With this information, the receiver Bob knows that $p_i$ lies on
a sphere of radius $r_i$ centred at $p_{i-1}$, but does not yet know its
precise direction to pinpoint its location on the sphere.  Assuming that $p_i$'s
direction is potentially uniformly distributed (consistent with a null
hypothesis), Alice can discretise the sphere's surface into cells (using a
common convention shared between Alice and Bob), each with area of
$\epsilon^2$. This discretisation can then be used to transmit $p_i$ as a cell
number $c_i$ over a variable length integer code.  With the knowledge of
$p_{i-1}$, $r_i$ and $c_i$, Bob can reconstruct $p_i$ to the stated  coordinate
accuracy of $\epsilon$ on each component. Thus, stating any  cell number takes
\begin{equation}
\label{eqn:codelen_uniform}
I_{\text{uniform}}(c_i) = -\log_2 \left(\dfrac{\epsilon^2}{4\pi r_i^2}\right) = \log_2 (4\pi r_i^2) -2\log_2
\epsilon~~\mathrm{bits}.
\end{equation}

Combining \myeqnspair{eqn:codelen_normal}{eqn:codelen_uniform}, each coordinate $p_i$ with respect to its previous coordinate $p_{i-1}$ can be stated raw using this sphere approach in
\begin{equation}
\label{eqn:codelen_sphere}
I_{\text{sphere}}(p_i) = I_{\mathcal{N}}(r_i) + I_{\text{uniform}}(c_i)~~\mathrm{bits}.
\end{equation}

Building on the above, a chain of coordinates of any protein
$\mathcal{P}\{p_1, p_2, \dotsc, p_{_{|\mathcal{P}|}}\}$ can be transmitted as a
null model message incrementally using this sphere approach.\footnote{Since the
original frame of reference of protein coordinates have no special meaning, the
coordinates can be translated such that the first coordinate $p_1$ lies on the
origin and, hence, does not need to be transmitted explicitly as part of the
message. Note, however, that even if $p_1$ is not assumed to be  the origin,
its encoding will add a constant cost to the total message length.} Alice sends
Bob the length of the protein ($|\mathcal{P}|$) over a variable length integer
code, followed by transmitting incrementally the coordinates $p_2$ given $p_1$,
then $p_3$ given $p_2$ and so on.   We note that the code length of
transmitting any positive integer $n$ over a variable length $\log^*$
distribution is:
\begin{equation}
\label{eqn:codelen_log*}
I_{\log^*}(n) = \log^*_2(n)+\log_2(2.865) \quad \text{bits}
\end{equation} 
where $\log_2^*(n) = \log_2 n + \log_2\log_2 n + \dotsb$ (over all +ve terms).

Therefore, the null model message length to describe any protein chain takes
\begin{equation}
I_{\text{null}}(p_1,\dotsc, p_{_{|\mathcal{P}|}}) = I_{\log^*}(|\mathcal{P}|) + \sum_{i=1}^{|\mathcal{P}|} I_{\text{sphere}}(p_i)
\quad\mathrm{bits}.
\label{eqn:null_chain_msglen}
\end{equation}

\section{Optimal encoding of a particular protein structure given a dictionary
for fragments} \label{sec:desc_protein} Let us assume that there exists a
dictionary of fragments $\mathcal{D}$ which both Alice and Bob agree to use to
communicate any chain of protein coordinates $\mathcal{P}=\{p_1,\cdots,
p_{_{|\mathcal{P}|}}\}$.

To encode the coordinates (i.e., compress the coordinates of \calP\ using
\calD), Alice's will dissect the protein chain into non-overlapping
regions, each of which is assigned to some dictionary fragment. These assigned
dictionary fragments are used as the basis to communicate the observed
coordinates in the respective regions of the protein economically. This is possible if only if the assigned dictionary
fragment has a good spatial fit with the region. 

Implicit to the model of communication, in addition to the dictionary, is the null model. For any regions in the protein, the null model competes with 
various possible explanations using the dictionary fragments -- 
if there is no compression
to be gained  (with respect to the null model message) using any one of 
the dictionary fragment to communicate some region, then the regions must be
transmitted raw.

Thus, this leads to an interesting optimization problem: find the best
dissection of proteins into (variable-length, non-overlapping) regions, each assigned to one of the fragments in the given dictionary or the null model. Before, we solve this problem \textit{optimally} (see \mysection{sec:optimaldissection}), we establish the procedure to encode the coordinates of a region using one of: (a) dictionary fragment (\mysection{sec:dictseg}) or (b) null model (\mysection{sec:coilseg}).

\subsection{Describing a region using a dictionary fragment}
\label{sec:dictseg} The methodology to communicate of any region $\{p_i,
\cdots, p_j\}\in \mathcal{P}, 1\le i<j\le (|\mathcal{P}|\equiv n)$ using an
assigned dictionary fragment $\mathcal{Q}=\{q_1,\cdots,q_{_{|\mathcal{Q}|}}\}\in\mathcal{D}$ is described here.
(Note, that a fragment can only be assigned provided the length of
$\mathcal{Q}$ is same as the length of the observed region, that is,
$|\mathcal{Q}| = j-i+1$ residues.)

Broadly, the coordinates of the region are transmitted in two steps. First, the
end point of the regions $p_j$ is transmitted, followed by each of its interior
points, $p_{i+1},p_{i+2},\cdots, p_{j-1}$.\footnote{Notice that the start point
$p_i$ is not transmitted -- the start point of the current region is also the
end point of the previous region, which the receiver would already knows,
assuming we are somewhere in between transmitting coordinates of \calP\ over
non-overlapping segments.}

\noindent\textit{Transmitting $p_j$.}
The end point $p_j$ is transmitted using the sphere
approach similar to the one described in \mysection{sec:nullmodel}.
The distance $d_{ij}$ between the start ($p_i$) and end ($p_j$) points is 
first transmitted using a normal distribution where the mean ($\mu$) is taken as
the distance $d^*$ between the start ($q_1$) and end points ($q_{_{|\mathcal{Q}|}}$) of the assigned fragment. The standard deviation $\sigma$ of the end point
is set to $\min\left((j-i)\times 0.2\text{\AA}, 3\text{\AA}\right)$ based
on the length of the fragment, $|\mathcal{Q}|$. 
Thus, using \myeqn{eqn:codelen_sphere}, stating the radius and direction using the the sphere approach takes $I_{\text{sphere}}(p_j)$ bits.

\noindent\textit{Transmitting interior points $p_{i+1},\cdots,p_{j-1}$}.  The
regions start and end points are now known to Bob, and there are $j-i-1$
interior points yet to be transmitted. Alice uses the following procedure to
transmit the interior points using an assigned dictionary fragment.  The
coordinates in \calQ\ are orthogonally transformed (i.e., rotated and
translated) to $\mathcal{Q}^\prime =\{q^\prime_1, q^\prime_2, \cdots,
q^\prime_{_{|\mathcal{Q}|}}\}$ such that: [a] $q^\prime_1$ is same as the start
point $p_i$ of the region, [b] the direction cosines of the vector connecting
the start and end points of the dictionary model,
$q^\prime_{_{|\mathcal{Q}|}}-q^\prime_1$, and the direction cosines of the
vector connecting the start and end points of the observed region, $p_j-p_i$,
are the same, and [c] the sum of the squared error of the interior points of
the region with the corresponding interior points of the dictionary model is
minimized.

This spatial transformation is related to the more general superposition
problem that minimizes the sum of the squared distance between two
corresponding vector sets which has an analytical solution~\cite{kearsley89}.
However, the transformation here is constrained such that the first points of
the two sets are the same (constraint [a]) and the rotational axis for the
dictionary model is the vector between the start ($p_i$) and end ($p_j$) points
of the region (constraint [b]). The first two constraints can be achieved using
elementary translation and rotation of the dictionary coordinates.  Thus
transformed,  the best rotation $\theta^*$ of $\mathcal{Q}$ about the $p_j-p_i$
axis is found so that constraint [c] is realized; this constrained version of
the problem can also be solved analytically using straightforward linear
algebra.

Once the transformation of \calQ\ to $\mathcal{Q}^\prime$ is achieved as
described above, Alice uses it to transmit the interior points of the region,
$p_{i+1},\cdots, p_{j-1}$, by: (1) transmitting the best rotation about the
$p_j-p_i$ axis so that Bob, after receiving it, can orient the nominated
dictionary fragment in the same orientation as Alice, and (2)  transmitting the
spatial deviations of the interior points with respect to the dictionary
fragment.

Rotation $\theta^*$  is
transmitted over an uniform distribution on a circle whose radius
$r_{\theta^*}$ is  the farthest distance of an interior point within 
the dictionary fragment from its axis of rotation. 
Dividing the circumference of a circle of radius $r_{\theta^*}$ 
into arcs of length $\epsilon$ and 
stating the arc number in which the rotated
coordinate with the farthest radius to the axis falls gives the
code length of stating $\theta^*$:
\vspace{-0.2cm}
\begin{equation} 
\label{eqn:codelen_theta}
I(\theta^*) = -\log_2 \left(\dfrac{\epsilon}{2\pi r_{\theta^*}}\right)
= \log_2 (2\pi r_{\theta^*}) -\log_2 \epsilon\quad\mathrm{bits}.
\end{equation}

Let the deviations  of interior coordinates with respect to
the corresponding (reoriented) dictionary fragment coordinates be denoted as $(\delta x_k,\delta y_k, \delta z_k)$. These components are transmitted using a normal distribution with a $\mu$ of 0 and
a standard deviation $\sigma$ set to the
sample standard deviation computed from these error components. (Note, $\sigma$ is the root-mean-squared distance between corresponding coordinates after the aforementioned orthogonal transformation.)

The MML estimate to transmit a set of
independent data $(\delta x_1, \delta y_1, \delta z_1)$, $(\delta x_2, \delta
y_2, \delta z_2)$, $\cdots$ $(\delta x_{_{|\mathcal{Q}|-2}}, \delta y_{_{|\mathcal{Q}|-2}}, \delta
z_{_{|\mathcal{Q}|-2}})$ 
using a normal distribution is given by (see \cite{wallace2005}): 
\begin{eqnarray} 
\label{eqn:codelen_spatialdev}
I(\delta x,\delta y, \delta z) &=&
\dfrac{1}{2}(N-1) \log_2 \sigma^2 + \dfrac{N-1}{2}\\\nonumber
&+& \dfrac{N}{2}\log_2
\left(\dfrac{2\pi}{\epsilon^2}\right) + \dfrac{1}{2}\log_2
\left(2N^2\right) \\\nonumber
&+& \log_2\left(R_\sigma\right)+ 1 +\log_2 \kappa_1
~\mathrm{bits}.
\end{eqnarray}
where $N = 3(|\mathcal{Q}|-2)$ is the total number of component spatial deviations being transmitted, $R_\sigma$ gives the (\textit{prior}) range on 
$\log_2 \sigma$, and $\kappa_1\approx \dfrac{1}{12}$
denotes the constant corresponding to quantizing lattices  proposed
by Conway and Sloane~\cite{conwaySloane84}. Here we bound $\sigma$ at 3\AA\ or less,  because this is consistent with the empirically observed 
limits of utility of root-mean-squared-deviation measure to
estimate similarity between rigid-body protein fragments. 

Therefore, combining all the code lengths described so far, the code length
required to transmit coordinates of a region of a protein using an
assigned dictionary fragment is
\begin{eqnarray} 
\label{eqn:codlen_segment}
I_\text{dict\_frag}(p_i,\cdots, p_j) = I_{\text{sphere}}(p_j) +
I(\theta^*) + I(\delta x, \delta y, \delta z) 
\end{eqnarray}

\subsection{Describing a protein region using the null model}
\label{sec:coilseg}
For encoding any region within a protein structure, there is an implicit null
model which any dictionary fragment encoding has to beat in order to be chosen. This ensures that incompressible regions (or \textit{random coils}) in proteins  can be stated, without the support of any dictionary fragment. Using \myeqn{eqn:null_chain_msglen}, the code length to transmit a segment of a protein  $p_i,\cdots, p_j$ as a random coil takes:
\begin{equation}
\label{eqn:codelen_coil}
I_{\text{coil}}(p_i,\cdots, p_j) = I_{\text{null}}(p_i,\cdots,p_j)
\quad\mathrm{bits}.
\end{equation}

\subsection{Describing a protein structure as a dissection of assigned
dictionary fragments} Here we describe the general language of communicating a
protein \calP\ given a dissection over a dictionary
$\mathcal{D}=\{\mathcal{Q}_1,\mathcal{Q}_2,\cdots,\mathcal{Q}_{_{|\mathcal{Q}|}}\}$
of fragments. (The search for optimal dissection is dealt in
\mysection{sec:optimaldissection}.)

Formally, a dissection of $\mathcal{P}=\{p_1,\cdots, p_n\}$ gives an ordered
subset of coordinates $\mathcal{P}^\prime = \{p^\prime_1\equiv p_{i_1},
p^\prime_2\equiv p_{i_2},\cdots,p_m^\prime\equiv p_{i_m}\}$ where
$1=i_1<i_2<\cdots<i_m=n$.  Each successive pair of coordinates in
$\mathcal{P}^\prime$, $\left<p^\prime_1,p^\prime_2\right>$,
$\left<p^\prime_2,p^\prime_3\right>$, $\cdots$
$\left<p^\prime_{m-1},p^\prime_m\right>$, defines a pair of start and end
points of a segment (or region). (Notice that the dissection
$\mathcal{P}^\prime$ gives $m-1$ successive, non-overlapping regions of \calP,
where end point of one segment is same as the start point of the next.)
Assigned to each region $\left<p^\prime_k,p^\prime_{k+1}\right>$ containing
$l_k=i_{k+1}-i_{k}+1$ points is an integer $t_k$ that specifies the model type
assigned to that region in the protein. This integer ranges from
$[1,|\mathcal{D}|+1]$, accounting for possible assignments over any one of
$|\mathcal{Q}|$ available fragments plus the implicit random coil (null) model.

Given some dissection of \calP\ using the dictionary \calD, Alice can transmit
the coordinates over a two part message: \noindent\textit{First part of the
message:} In the first part of the message, Alice communicates the
segmentation, and the corresponding assignments of models (dictionary fragment
or coil) as a hypothesis on the observed coordinate data.  To achieve this, the
following information is transmitted. Using \myeqn{eqn:codelen_log*}, stating
the number of segments ($m-1$) in the dissection takes $I_{\log^*}(m-1)$ bits.
The segmentation can  be stated simply as a sequence of assigned model types.
Assuming each model type is uniformly likely, each integer $t_k$ corresponding
to the assigned model type  takes 
\begin{equation}
\label{eqn:codelen_uniform2}
I_{\text{uniform}}(t_k) = \log_2 (|\mathcal{D}|+1) \quad \mathrm{bits}.
\end{equation}

Thus, the message length of the first part requires
\begin{equation}
\label{eqn:codelen_first}
I_\text{first}(\mathcal{P}^\prime|\mathcal{D}) = I_{\log^*}(m-1) 
+ \sum_{k=1}^{m-1} I_{\text{uniform}}(t_k)   \quad \mathrm{bits}
\end{equation}

\noindent\textit{Second part of the message:} In the second part, Alice
transmits  the coordinates of \calP\ using the hypothesis described in the
first part of the message. The procedure to transmit coordinate data of a
segment of a protein using dictionary fragment and coil models have already
been discussed in sections \ref{sec:dictseg} and \ref{sec:coilseg}.

 The message length of the second part is given by: 
\begin{eqnarray}
\label{eqn:codelen_second}
I_{\text{second}}(\mathcal{P}|\mathcal{P}^\prime\&\mathcal{D}) 
= \sum_{k=1}^{m-1} I_{\text{model}}(p^\prime_k,\cdots, p^\prime_{k+1})
\quad\text{bits}
\end{eqnarray} where  $I_{\text{model}}(p^\prime_k,\cdots, p^\prime_{k+1}) =
I_{\text{dict\_frag}}(p^\prime_k,\cdots, p^\prime_{k+1})$ if $t_k$ corresponds
to a dictionary fragment and $I_{\text{model}} =
I_{\text{coil}}(p^\prime_k,\cdots, p^\prime_{k+1})$  if $t_k$ corresponds to a
random coil.

Thus, the total message length of communicating the coordinates of a particular
protein $\mathcal{P}$ using some dissection $\mathcal{P}^\prime$ over a given
dictionary \calD\ is computed by combining the terms in
\myeqnspair{eqn:codelen_first}{eqn:codelen_second}:
\begin{eqnarray}
\label{eqn:codelen_protein}
I(\mathcal{P}^\prime\&\mathcal{P}|\mathcal{D}) =
I_{\text{first}}(\mathcal{P}^\prime|\mathcal{D}) +
I_{\text{second}}(\mathcal{P}|\mathcal{P}^\prime\&\mathcal{D}) 
\end{eqnarray}

\subsection{Search for the optimal dissection of a particular protein given a
\label{sec:optimaldissection}
dictionary of substructures} Given \myeqn{eqn:codelen_protein}, the problem of
inferring the best dissection can now be stated formally as follows: Given a
protein $\mathcal{P}$  and a dictionary \calD, find the dissection
$\mathcal{P}^\prime$ of \calP\ such that the message length to transmit
$\mathcal{P}$ losslessly is \textit{minimized}.

In the protein \calP, any pair of points can  potentially be the start and end
points of a dissected segment. At the same time, any dissected segment can
potentially be described using a coil model or any of the dictionary fragments
(provided the length of the assigned dictionary element is same as the length
of the segment).  Therefore, the procedure to assign dictionary fragments to
coordinates $\{p_1,\cdots, p_{_{|\mathcal{P}|}}\}$  in  a given protein begins
by constructing a set of $|\mathcal{D}|+1$ code length matrices, one for each
possible encoding model type $t$. Using
\myeqnspair{eqn:codelen_first}{eqn:codelen_second}, the code length matrices
are filled as: \begin{eqnarray} H^{t}(i,j) = I_{\log^*}(j-i) +
I_{\text{uniform}}(t)+I_{\text{model}}(p_i,\cdots, p_j) \end{eqnarray} where
any cell $(i,j)_{_{i\le i,j\le |\mathcal{P}|}}$ of the matrix for type $t$
gives the code length of stating the segment $p_i$ to $p_j$ using the model
$t$.

\noindent\textit{Optimal dissection as a dynamic program over the code length
matrices, $\left\{H^1, \cdots H^{^{|\mathcal{D}|}}, H^{^{|\mathcal{D}|+1}}\right\}$}: The dissection of $\mathcal{P}$ using various model types
enforces a linear ordering constraint with optimal substructure and overlapping
subproblems which can be composed incrementally.  This allows the search for an
optimal dissection of a protein structure, \textit{given} a dictionary, to be
formulated using a dynamic program that can be solved efficiently.

Let any $M(i)$ store the optimal encoding message length to 
transmit points $p_1,\cdots, p_i$, for all $1\le i\le |\mathcal{P}|$. With the
boundary condition of $M(1) = 0$, the dynamic programming recurrence 
to find the optimal assignment is given
by 
\begin{equation*}
\displaystyle
M(j) = \min_{i=1}^{j-1}\left\{
\begin{array}{l}
\displaystyle\min_{t=1}^{|\mathcal{D}|+1} H^t(1,j) 
\\~\\
M(i) + \displaystyle\min_t H^t(i,j)
\end{array}\right.
{\scriptsize t\implies \text{model}=
  \left\{
  \begin{array}{l}
    \mathcal{Q}_t\in\mathcal D\\
    \text{coil}
  \end{array}
  \right.} 
\end{equation*}

The above recurrence is used to fill the array $M$ iteratively from
$1$ to $|\mathcal{P}|$. On completion, the best assignment of dictionary fragments 
can be derived by remembering the index $i$ and type $t$ from which the
optimal $M(j)$ is computed. The time complexity to find the optimal encoding is $O(|\mathcal{D}||\mathcal{P}|^2)$. 

\section{Describing an entire collection of protein structures using a given
dictionary of substructures} \label{sec:describecollection} Restating
\myeqn{eqn:mainobjective} in \mysection{sec:dictmml},
$I(\mathcal{D}\&\mathcal{C}) = I(\mathcal{D}) + I(\mathcal{C}|\mathcal{D}))$.
That is,the \textit{optimal encoding of a collection of protein structures}
\calC\ using a given dictionary \calD\ involves a one-off cost of stating the
dictionary ($I(\mathcal{D})$), followed by the optimal encodings of the
individual proteins in the collection ($I(\mathcal{C}|\mathcal{D}))$.  Below we
show the details of how the respective terms are estimated.
\subsection{Estimating $I(\mathcal{D})$ for a given dictionary} For the
dictionary $\mathcal{D}$  with substructures
$\{\mathcal{Q}_1,\mathcal{Q}_2\cdots,\mathcal{Q}_{_{|\mathcal{D}|}}\}$, the
statement cost, $I(\mathcal{D})$, involves:  the transmission of the the number
of substructures ($|\mathcal{D}|$), followed by  the statement cost of each
fragment's coordinates $\mathcal{Q}_i\in\mathcal{D}$ over a null model message.

From Equations \ref{eqn:codelen_log*} and \ref{eqn:null_chain_msglen}, we can
derive the message length of stating the dictionary \calD\ as: 
\begin{equation}
I(\mathcal{D}) = I_{\log^*}(|\mathcal{D}|)+ \sum_{i=1}^{|\mathcal{D}|} I_{\text{null}}(\mathcal{Q}_i)
\quad\mathrm{bits}. 
\label{eqn:dictcost}
\end{equation} 

\subsection{Estimating $I(\mathcal{C}|\mathcal{D})$}
\label{sec:objective}
With the dictionary transmitted to Bob, Alice can now encode the structures 
in the collection using the method described in Section \ref{sec:desc_protein}.
Using \myeqn{eqn:codelen_protein}, the statement cost to transmit the coordinates of a collection of proteins $\mathcal{C} = \{\mathcal{P}_1, \cdots,
\mathcal{P}_{_{|\mathcal{C}|}}\}$ using the previously stated dictionary $\mathcal{D}$ takes:

\begin{equation}
I(\mathcal{C}|\mathcal{D}) = \sum_{i=1}^{|\mathcal{C}|}
I(\mathcal{P}_i^\prime\&\mathcal{P}_i|\mathcal{D})) \quad \text{bits}
\label{eqn:collection}
\end{equation}

\section{The search for the optimal dictionary}
\label{sec:searchdictionary}
\mysection{sec:objective} has established a rigorous objective to search for the dictionary of fragments building blocks of protein three-dimensional structures. It follows from the aforementioned objective (also \myeqn{eqn:mainobjective}) that an optimal dictionary for a collection of protein structures is one for which the statement of the dictionary, plus the sum of the optimal encodings of all the proteins in the set, is the \textit{shortest}. That is, the objective of this work is to find a $\mathcal{D}^*$ such that: 

\begin{equation}
I(\mathcal{D}^*\&\mathcal{C}) = \min_{\forall\mathcal{D}} 
I(\mathcal{D} \&\mathcal{C})
\quad \text{bits}.
\label{eqn:finalobjective}
\end{equation}

Clearly, any fragment (of arbitrary size) from within any protein in the
collection is a potential candidate for the dictionary. Therefore searching for
the best dictionary leads to a very large optimization problem. Since the
problem is computational intractable  to find a provably optimal dictionary, we
designed a simulated annealing algorithm in order to evolve a dictionary that
iteratively converges to the best dictionary defined by the 
\myeqn{eqn:finalobjective}.

Simulated Annealing is an heuristic approach which has an analogy with cooling
of solids. Here, we consider each possible dictionary (of arbitrary number of
fragments) as being analogous to some state of a physical system. The message
length of transmitting a collection of structures using any dictionary given by
Equation \ref{eqn:mainobjective} is analogous to the internal energy of the
physical system in that state.

The method starts with an empty dictionary. In this state, each protein in the
collection is transmitted raw, as a random coil using the null model. The
strategy involves iteratively perturbing the dictionary from this 
initial empty state to a state where the total message length
objective is minimized. 

\begin{figure*}[h]
\begin{center}
\begin{tabular}{cccc}
\includegraphics[width=42mm]{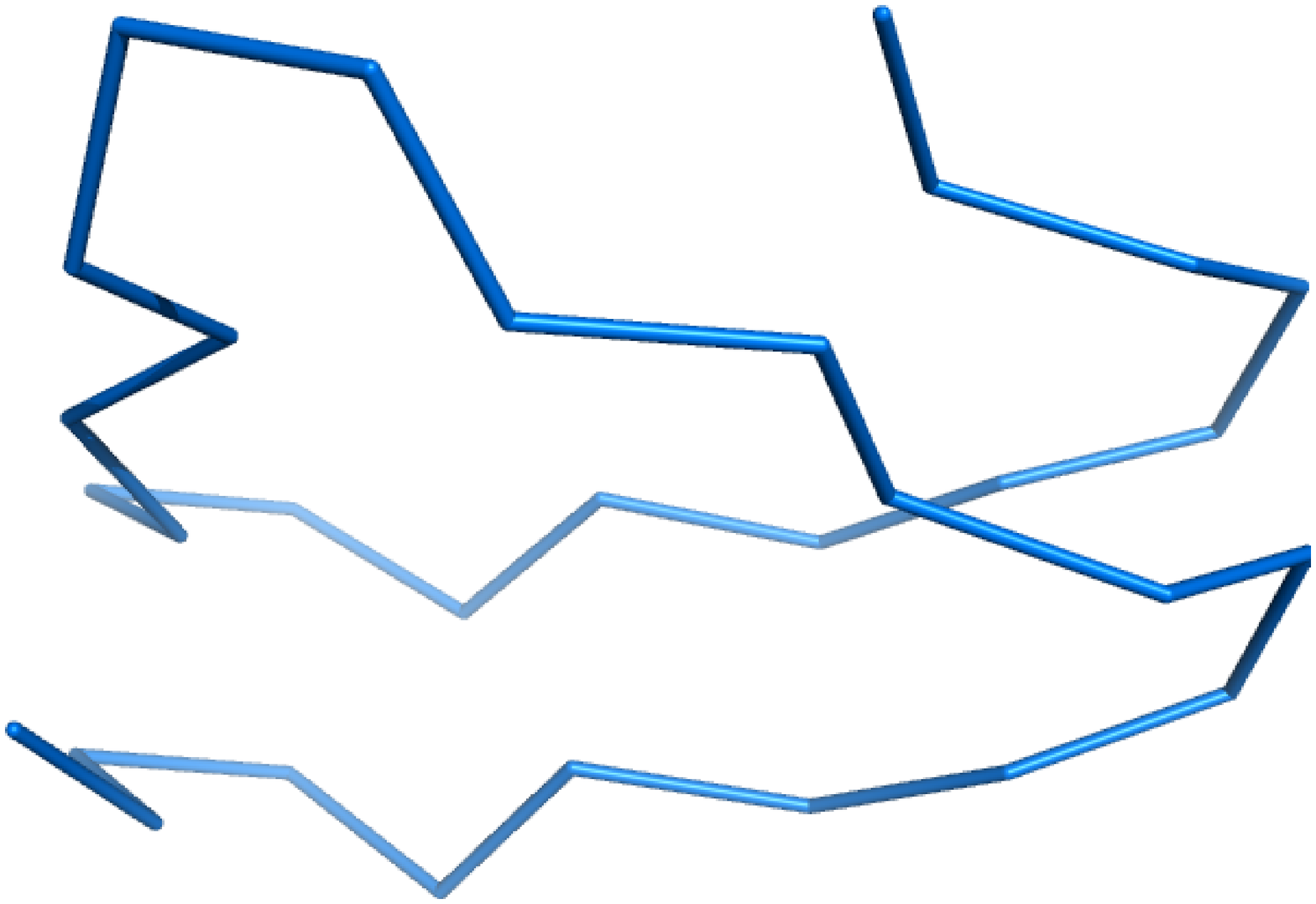} &
\includegraphics[width=42mm]{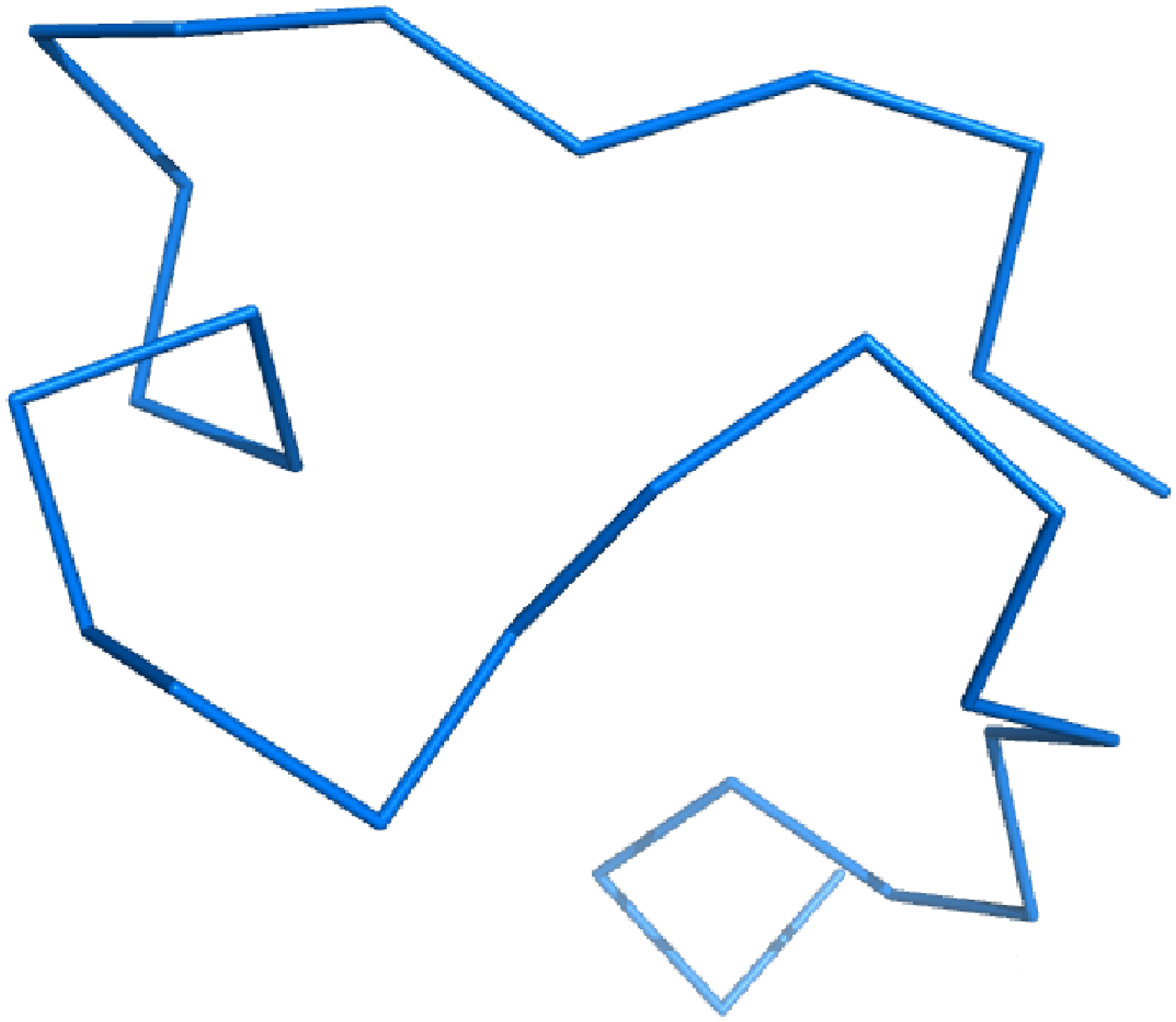} &
\includegraphics[width=42mm]{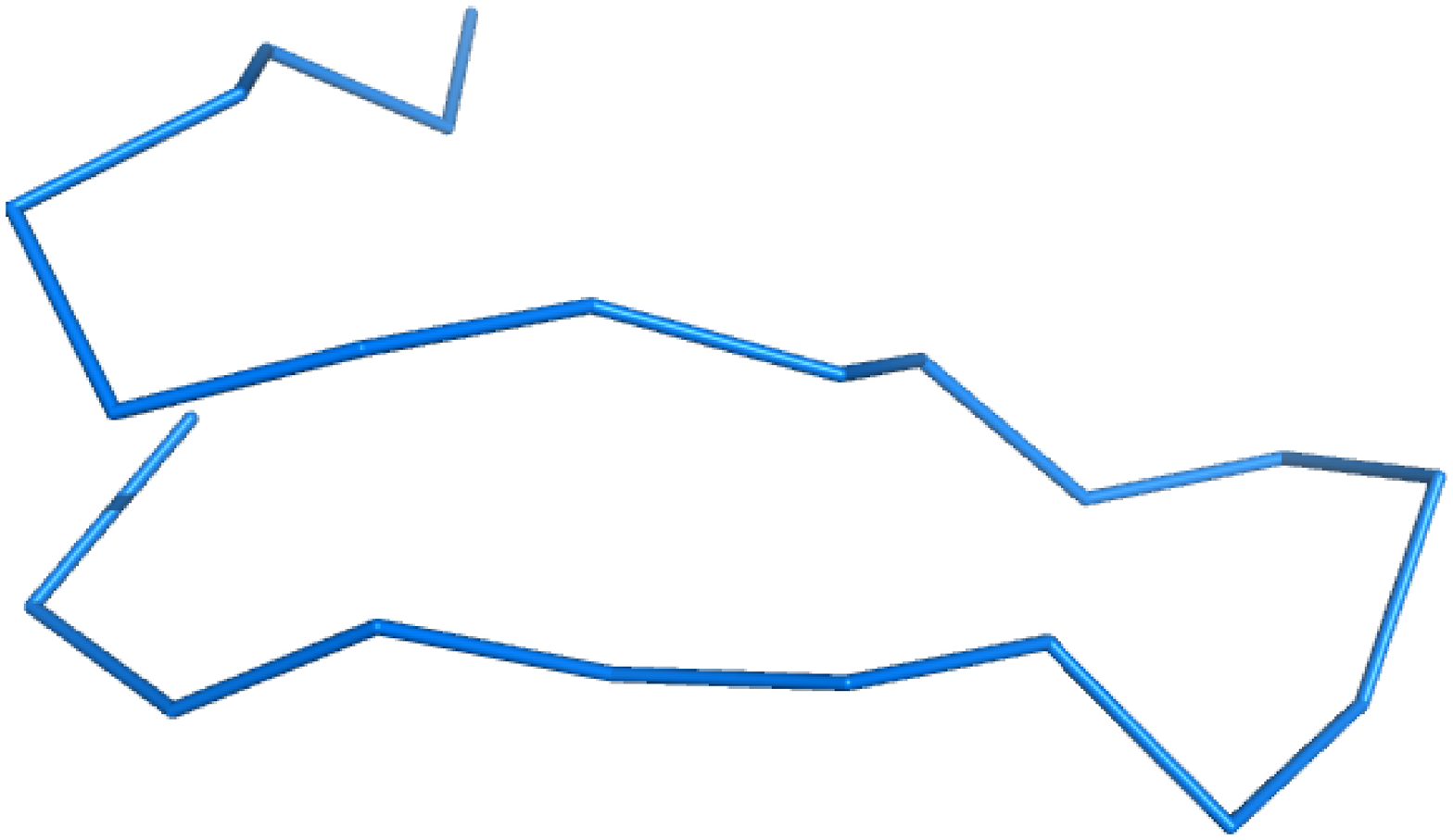} &
\includegraphics[width=42mm]{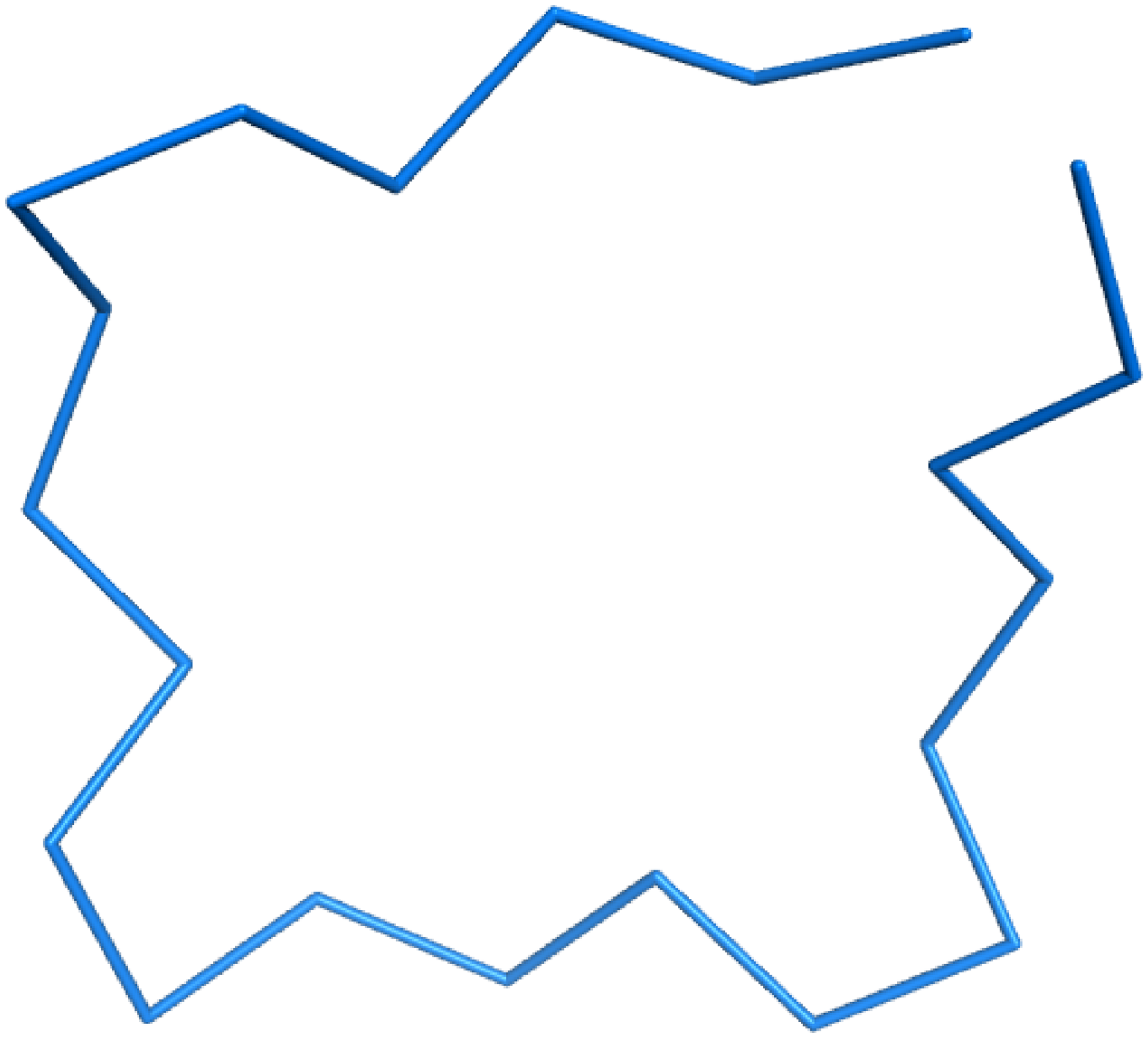} \\
(a) & (b) & (c) & (d)\\
\end{tabular}  
\end{center}
\caption{Four fragments chosen from our dictionary our approach discovered. (a)  $1\frac{2}{3}$ turns of three-fold $\beta$-helix of length 31 residues. It occurs 26 times in our collection.  See wwPDB 2IC7 for one such instance. (b) An exotic $\beta$-hairpin of length 29 residues. This occurs 20 times in the collection. See wwPDB 1UJU for one such instance. (c) A long $\beta$-hairpin of length 22 residues with 47 occurrences. See wwPDB 1JO8 for one such instance. (d) 1 turn of four-fold $\beta$-helix of length 21, which occurs 19 times in the collection.}
\label{fig:piccies}
\end{figure*}

\begin{figure*}[h]
\begin{center}
\hspace{-2in}
\begin{minipage}[t]{45mm}
\vspace{0pt}
\includegraphics[width=42mm]{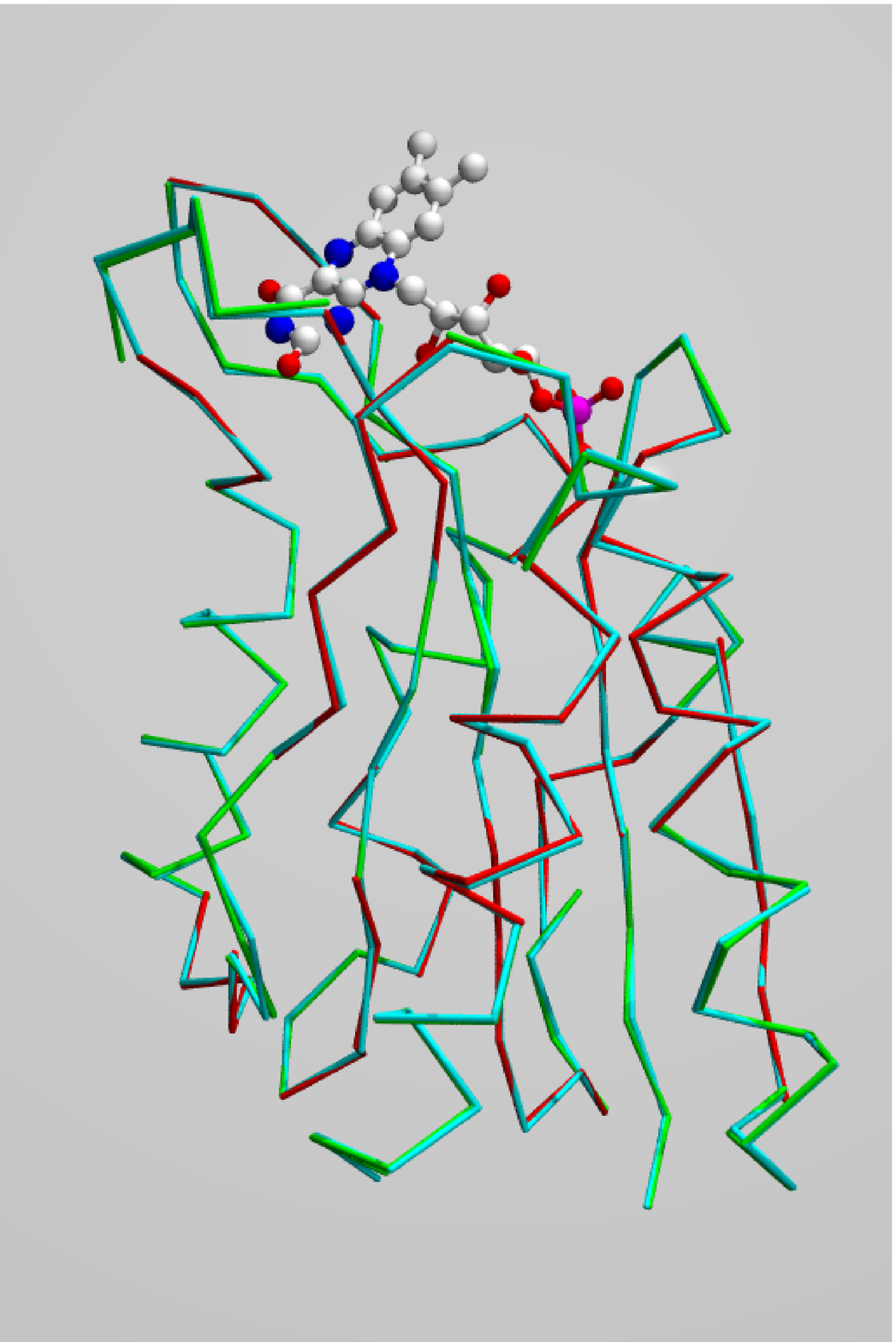} 
\end{minipage}%
\begin{minipage}[t]{45mm}
\vspace{0pt}
\begin{tabular}{|rl@{~}c@{~}||@{~}rl@{~}c|}
\hline
Resi&Model& RMSD (\AA)& Resi&Model& RMSD (\AA)\\\hline
2  - 6  &  m1096 &  0.12 &  78  -  82 &  m1415 &  0.14  \\
6  - 10 &  m1195 &  0.21 &  82  -  85 &  m1623 &  0.07  \\
10 - 13 &  m1595 &  0.24 &  85  -  89 &  m1083 &  0.20  \\
13 - 22 &  m0231 &  0.16 &  89  -  92 &  m1706 &  0.03  \\
22 - 32 &  m0159 &  0.18 &  92  -  95 &  m1611 &  0.06  \\
32 - 37 &  m0874 &  0.27 &  95   -  100 &  m0967 &  0.53\\ 
37 - 41 &  m1202 &  0.24 &  100  -  103 &  m1571 &  0.17\\
41 - 45 &  m1246 &  0.18 &  103  -  115 &  m0054 &  0.12\\
45 - 49 &  m1323 &  0.17 &  115  -  119 &  m1499 &  0.19\\
49 - 55 &  m0502 &  0.19 &  119  -  123 &  m1480 &  0.21\\
55 - 60 &  m0930 &  0.28 &  123  -  128 &  m0769 &  0.29\\
60 - 63 &  m1685 &  0.07 &  128  -  133 &  m0750 &  0.32\\
63 - 67 &  m1306 &  0.26 &  133  -  141 &  m0281 &  0.12\\
67 - 71 &  m1194 &  0.16 &  141  -  148 &  m0426 &  0.19\\
71 - 78 &  m0423 &  0.12 &              &        &      \\\hline
\end{tabular}
\end{minipage}
\end{center}

\caption{Dissection of the structure of
flavodoxin from \textit{Desulfovibrio vulgaris} (wwPDB entry 1J8Q).
Parent structure shown in cyan; successive dictionary fragments
in alternating red and green.  Cofactor flavin mononucleotide (FMN)
in ball-and-stick representation. Dissected regions are listed as a table below the picture.}
\label{fig:flavodoxin}
\end{figure*}

\noindent\textit{Perturbations} At each step the current dictionary is
perturbed randomly (which is akin to sampling some new nearby dictionary
state). 

The choice of moving to the new state or remaining in the current one is
decided probabilistically. Specifically, at each iteration, our method employs one of the following randomly chosen perturbations: 

\noindent{\texttt{Add:}} Append to the current dictionary a new fragment. This fragment is from a randomly chosen protein from the collection, of random length.\\

\noindent{\texttt{Remove:}} Remove a randomly chosen fragment from the existing dictionary state.\\

\noindent{\texttt{Swap:}} Replace a randomly chosen fragment in the current dictionary with another randomly chosen fragment from the collection. This is equivalent to the sequence of perturbations: `\texttt{Remove}' followed by an `\texttt{Add}'.\\

 \noindent{\texttt{Perturb length:}} Expand or shrink a randomly chosen fragment from the current dictionary state by one residue, at the randomly chosen end. (Expanding a previously chosen fragment is achieved by remembering its locus -- i.e., the fragment's source protein structure and its offset in the source. This allows any existing fragment to be elongated by an additional coordinate, based on the information available in its source structure.)

\textit{Probability of acceptance or rejection of any perturbation}. Equation \ref{eqn:mainobjective} gives the estimate of the negative logarithm of the joint probability of a dictionary and a collection. 
In \mysection{sec:mml} we have seen that  the difference between the 
message lengths using any two different hypotheses 
(here, two dictionary states) gives the log-odds posterior ratio. 
This implies, if the total message length using a perturbed
dictionary is $k$ bits shorter (conversely, longer) than the current state, 
then the perturbed  dictionary is $2^{k}$ times more likely (conversely, 
unlikely) than the perturbed state.  

The simulated annealing heuristic  starts with a high (time-varying)
parameter $t$ (after temperature). Let $I_{\text{current}}$ and
$I_{\text{perturbed}}$ be the total message lengths using the 
current and perturbed states of a dictionary. During any
iteration, if $\Delta I \equiv (I_{\text{current}} - I_{\text{perturbed}}) < 0$, the perturbed state is immediately accepted as the new current state, and the procedure continued Otherwise, the perturbed state is accepted with a probability of 
$1/{2^{^{\scriptsize \frac{\Delta I}{t}}}}$.

\noindent\textit{Cooling schedule}.
For simulated annealing algorithms, the variation of the 
temperature parameter $T$ controls the evolution of 
the dictionary states.
We set $T$ to a high $10,000$ at the start. The parameter $T$ 
decays at a constant rate of 0.88. For each value of $t$ between $10,000$ and  $10$,  we carry out $10,000$ perturbations, while for
value of  $T$ less than 10, we carry out  $100,000$ perturbations to the dictionary. 
The stopping criterion is when the number of iterations reaches 2 million iterations (which occurs at $t=0.246$).

\noindent\textit{Implementation:} A program to optimally encode a given collection of
structures and then search for the best dictionary using simulated annealing
was implemented in the \texttt{C++} programming language. Message Passing
Interface (\texttt{MPI}) was used to parallelize this program  to run on a
large computational cluster.
\ignore{(\url{www.massive.org.au}) located at Monash
University and Australian Synchrotron.}

A MapReduce model was used to distribute the encoding tasks on the large
cluster. In this model, a master node accepts a collection of protein
structures as input and evenly divides them into smaller subsets of structures,
where each subset is distributed to worker nodes. For every perturbation of a
current dictionary state, each worker node computes the sum of message lengths
to optimally encode each of the structures in its allocated subset.  The worker
node then  returns the message length back to the master node, which collects
all the message lengths and combines them to compute the total message length
to encode all the structures in the given collection using the current state of
the dictionary. 

\section{Results} \label{sec:results} The work resulted in a dictionary of 1711
fragments ranging in length from 4 to 31 amino acids. This dictionary was used
to dissect the entire collection of 8992 source structures.  The average
root-mean-square (r.m.s.) deviation of orthogonal superposition of the
dictionary fragments to the assigned regions is 0.29 {\AA}. (Note, this does not
include the separate statement and application of deviations which is part of
the transmitted message, which would reduce 0.29 {\AA} to 0.)  The average,
over all proteins in the set, of the maximum r.m.s.~deviation of any model from
all regions it encodes, is 1.23 {\AA}. In \myfig{fig:piccies} shows the visualization of four fragments chosen from the dictionary we discovered, of lengths 31, 29, 22 and 21 respectively. Previous methods, due to the length constraint (see \mysection{sec:intro}), are unable to detect recurrent fragments this long.

 Figure
\ref{fig:flavodoxin} shows the optimal dissection into dictionary fragments of the structure of flavodoxin
from \textit{Desulfovibrio vulgaris} (wwPDB entry 1J8Q, solved at 1.35
{\AA} 
resolution~\cite{flavodoxin}.)
To encode losslessly 
the flavodoxin structure, the dissection would be accompanied
by the vector 
deviations of the C$_\alpha$ atoms of flavodoxin from the
assigned canonical dictionary fragments.
Noteworthy properties of the dissection in this example include:

\begin{enumerate}
\item The fits of the individual fragments of the dictionary to the structure
are quite precise in almost all cases.  The deviations are of the order
of only tenths of an {\AA}.  (The maximum r.m.s.d. between the coordinates of
a region in the protein and the assigned model is 0.53 {\AA}; this occurs only
once. All other r.m.s.d.~values are $\leq$ 0.32 {\AA}.)
\item Here the dictionary fragments account for the entire structure.
No regions of the structure need be encoded as a random coil; that is, using the
null model.
\item The range of lengths of the dictionary fragments 
appearing in the dissection
of flavodoxin is from 4 to 13.
Some of the segments
correspond to individual fragments of secondary structure -- helices and strands of sheets.  Others correspond to N- or C-terminal parts of
secondary
structures, plus parts of the loops either preceding or succeeding
them.
\item The sequence of dictionary fragments
provides a one-dimensional representation of the protein 
folding pattern.
\end{enumerate}

\begin{table}[!t]
\renewcommand{\arraystretch}{1.3}
\caption{Clustering of models in the identified dictionary}
\label{table:clusters}
\centering
\begin{tabular}{|r@{~~}r@{~~}c@{~~}l|}
\hline
\textbf{Class}&\textbf{Size}&\textbf{Code}&\textbf{Description}\\\hline
1 & 616 & e & short extended regions \\
2 & 301 & t & short non-hairpin turns {(some $\beta-$bulges)}\\
3 & 4   & t & short non-hairpin turns \\
4 & 325 & h & short helices \\
5 & 164 & h & short helices\\
6 & 167 & E & extended regions  {(some with hooks at end)}\\
7 & 6   & E & extended regions  {(some curved)}\\
8 & 13  & T & non-hairpin turns {(some $\beta-$bulges)}\\
9 & 50  & b & shorter $\beta-$hairpin \\
10 & 30 & B & $\beta-$hairpins \\
11 & 3  & B & $\beta-$hairpins, unequal arms {(`shepherd's crook')}\\
12 & 1  & B & $\beta-$hairpins, unequal arms {(`shepherd's crook')}\\
13 & 3  & H & irregular alpha helix (plus C$_\alpha$-only)\\
14 & 17 & H & long alpha helices\\
15 & 2  & $\Omega$ & long wide loops { ($\Omega$ loop)}\\
16 & 1  & $\Omega$ & long wide loops { ($\Omega$ loop)}\\
17 & 2  & $\Omega$ & long wide loops { ($\Omega$ loop)}\\
18 & 1  & C & double $\beta-$hairpin { (`paper clip')}\\
19 & 1  & B & long twisted $\beta-$hairpins \\
20 & 2  & $\Omega$ &helix-strand-helix-strand / wide ($\Omega$) loops\\
21 & 1  & 3 & 1$\frac{2}{3}$ turns of three-fold $\beta-$helix \\
22 & 1  & 4 &1 turn of four-fold   {($\beta-$helix)} \\\hline
\end{tabular}
\end{table}

\noindent\textit{Clustering of the dictionary fragments:}
To further rationalize the 1711 dictionary fragments, 
we clustered the dictionary fragments into coarse structural classes with
the UPGMA method~\cite{upgma} using the Mahalanobis 
distance~\cite{mahanalobisdist} computed from the following characterising properties:\\
(1) the number of backbone hydrogen bonds between residues 
    separated by 4 in the 
    sequence (to group helices which demonstrate this periodicity),\\
(2) the distance between the C$_\alpha$ atoms of N- and C-terminal
residues,\\ 
(3) the cosine of the angle between the C$_\alpha$ atom of
the N-terminal residue, the C$_\alpha$ atom of the middle residue
(or, for fragments containing even numbers of residues,
the average position of the C$_\alpha$ atoms of the two middle
residues),\\
(4) the r.m.s. deviation of a fit of the C$_\alpha$ atoms to
a straight line, and\\ 
(5) the average value of the cosine of the dot
products of C$\rightarrow$O vectors of successive residues.

\mytab{table:clusters} in the main text shows the clusters, and suggested class codes.

Taking together the dissections of all proteins in the source set,
the usage of different models in terms of case-insensitive coarse classes defined in \mytab{table:clusters} in main text. The distribution
of usage, showing the prominence of secondary-structure elements, is:

\begin{center}
\begin{tabular}{p{2.7cm}p{2.7cm}p{2.7cm}p{2.7cm}p{2.7cm}p{2.7cm}p{2.7cm}p{2.7cm}}
\multicolumn{1}{c}{E}&\multicolumn{1}{c}{H}&\multicolumn{1}{c}{T}&\multicolumn{1}{c}{B}&\multicolumn{1}{c}{$\Omega$}&\multicolumn{1}{c}{C}&\multicolumn{1}{c}{3}&\multicolumn{1}{c}{4}\\
\multicolumn{1}{c}{180360}&\multicolumn{1}{c}{107262}&\multicolumn{1}{c}{72096}&\multicolumn{1}{c}{12390}&\multicolumn{1}{c}{447}&\multicolumn{1}{c}{48}&\multicolumn{1}{c}{28}&\multicolumn{1}{c}{21}\\
\end{tabular}
\end{center}

The distribution of two-letter combinations is:

\begin{center}
\begin{tabular}{|cr|cr|cr|cr|}
EE & 87644&HH & 44008&HE & 39951&EH & 38735\\
ET & 38401&TE & 36721&TH & 21153&HT & 18778\\
TT & 11583&EB & 9074&BE & 8797&BT & 1790\\
TB & 1460&HB & 1247&BH & 1223&BB & 473\\
EO & 306&OE & 232&OH & 169&TO & 70\\
HO & 62&EC & 48&CE & 35&OT & 23\\
3E & 22&OB & 18&E4 & 15&E3 & 15\\
4E & 13&CT & 12&T3 & 8&44 & 6\\
BO & 6&3T & 3&33 & 3&4T & 2\\
H3 & 1&CB & 1&&\\
\end{tabular}
\end{center}

\noindent 
Repetitions --  EE and HH -- lead the
list, and that symmetric pairs (HE/EH, and ET/TE) tend to
have similar distributions.  The first 8 pairs account for
90\% of the two-character combinations.  This confirms the
well-known
importance of secondary structures as building blocks of
proteins (assuming that most of the E regions are in fact
parts of $\beta-$sheets; conversely, many of the $\beta-$hairpins
include strands of sheet.)  The first combination that does not
contain any secondary structural element, TT (T = non-hairpin turn)
accounts for 3\% of the total number of pairwise combinations. 

\noindent\textit{One-dimensional representation of protein folds.}
We consider as a case study the assignment of
fragments to clusters which reveals structural patterns in 
Drosophila lebanonensis \textit{alcohol dehydrogenase} (wwPDB code 1SBY).

\begin{figure*}{h}
\centering

\begin{tabular}{|ll@{~~}c@{~~}l@{~}c@{~}c@{~}lc@{~}c|}
\hline
Model&Residues&Length& r.m.s.d.&Class&Code&Description of
class&Secondary&Residues\\
&&&&&&&structure&\\
\hline
m1685 & 1   - 4 & 4 & 0.168 & 1 & e & short extended region & & \\
m0546 & 4   - 10 & 7 & 0.234 & 6 & E & extended region & $\beta1$ & 7 - 11\\
m1700 & 10  - 13 & 4 & 0.056 & 2 & t & short turn & &\\
m1400 & 13  - 17 & 5 & 0.442 & 2 & t & short turn & &\\
m0046 & 17  - 29 & 13 & 0.498 & 4 & h & short helix & $\alpha$1 & 15 - 28\\
m1138 & 29  - 33 & 5 & 0.085 & 2 & t & short turn & &\\
m0602 & 33  - 39 & 7 & 0.538 & 6 & E & extended region & $\beta$2 & 32 - 37\\
m1556 & 39  - 42 & 4 & 0.136 & 1 & e & short extended region & & \\
m0195 & 42  - 52 & 11 & 0.174 & 4 & h & short helix & $\alpha$2 & 41 - 52\\
m0967 & 52  - 57 & 6 & 0.183 & 1 & e & short extended region & &\\
m0855 & 57  - 62 & 6 & 0.230 & 1 & e & short extended region & $\beta$3 & 56 - 61\\
m1247 & 62  - 66 & 5 & 0.170 & 2 & t & short turn & &\\
m0243 & 66  - 74 & 9 & 0.200 & 4 & h & short helix & &\\
m0316 & 74  - 82 & 9 & 0.142 & 4 & h & short helix & &\\
m0791 & 82  - 87 & 6 & 0.304 & 1 & e & short extended region & &\\
m1188 & 87  - 91 & 5 & 0.220 & 1 & e & short extended region & $\beta$4 & 87 - 90\\
m1291 & 91  - 95 & 5 & 0.299 & 1 & e & short extended region & &\\
m0804 & 95  - 100 & 6 & 0.226 & 2 & t & short turn & &\\
m0315 & 100 - 108 & 9 & 0.130 & 4 & h & short helix & &\\
m0096 & 108 - 119 & 12 & 0.182 & 4 & h & short helix & $\alpha$4 & 108 - 123\\
m1120 & 119 - 123 & 5 & 0.168 & 2 & t & short turn & &\\
m0600 & 123 - 129 & 7 & 0.257 & 2 & t & short turn & &\\
m0668 & 129 - 134 & 6 & 0.413 & 1 & e & short extended region & $\beta$5 & 131 - 136\\
m1083 & 134 - 138 & 5 & 0.215 & 1 & e & short extended region & &\\
m1243 & 138 - 142 & 5 & 0.140 & 5 & h & short helix & &\\
m0408 & 142 - 149 & 8 & 0.660 & 2 & t & short turn & &\\
m0177 & 149 - 159 & 11 & 0.160 & 4 & h & short helix & $\alpha$5 & 148 - 173\\
m0180 & 159 - 169 & 11 & 0.165 & 4 & h & short helix & &\\
m1260 & 169 - 173 & 5 & 0.151 & 2 & t & short turn & &\\
m0636 & 173 - 179 & 7 & 0.448 & 6 & E & extended region & $\beta$6 & 174 - 171\\
m1188 & 179 - 183 & 5 & 0.152 & 1 & e & short extended region & &\\\hline
\end{tabular}

\caption{Dissection of alcohol dehydrogenase structure (1SBY) 
  using dictionary fragments.}
\label{fig:1sby}
\end{figure*}

Figure \ref{fig:1sby} shows
the dissection of amino acid residues 1-183 of 1SBY
into dictionary fragments. In addition
to our
dissection, we show the
secondary structure assignments
from the wwPDB file for the NAD-binding domain
supersecondary structure.

The sequence of class symbols (column 6 in table in 
Figure \ref{fig:1sby}),
converted to upper case to suppress the 
distinction between
short and long versions of the same substructure, affords a
more perspicuous representation of this dissection:

\begin{center}
\texttt{EETTHTEEHEETHHEEETHHTTEEHTHHTEE}
\end{center}
\noindent (E = strand, H = helix, T = non-hairpin turn (Table 1).)

\noindent This is an instance of the regular expression:
\begin{center}
\texttt{(E+T+H+T+E+HE*TH+T*E+)\{2\}}
\end{center}

This sequence from the dissection provides a concise
one-dimensional representation of the folding pattern.  It captures
the duplication of the two $\beta-\alpha-\beta-\alpha-\beta$ 
substructures (and the
points of insertion of non-pattern elements) but not their symmetrical
spatial disposition. Nevertheless, the dissection provides a faithful
signature of the NAD-binding domain folding pattern~\cite{lesk95}.

The rich repertoire of algorithms on character strings
is applicable.  For example, the string could be used to design
regular expressions for probing collections for similar structures.

More generally, standard regular-expression-matching 
algorithms permit application of the linear representations
to identify common folding patterns in a set of structures, or,
specifically, to classify a newly-determined structure,  as for example in SCOP~\cite{scop} and CATH~\cite{cath}.  The
representation can also identify variations and deviations from
standard folding patterns in known families.  
For instance, some
NAD-binding domains (including \textit{Drosophila lebanonensis}
alcohol dehydrogenase) contain extra helices and/or
hairpins~\cite{rose99} and this
would be revealed by a `sequence' alignment of the dissections of
members of this family.

\section{Conclusion}
This work introduces a novel method to infer the dictionary  of building blocks
of protein structures. This work fall squarely into one of the most important cross-disciplinary areas of modern science, where biology and computing meet.
The  approach described in this paper is a successful demonstration of rigorous statistical inference applied to an important data mining problem in structural Bioinformatics. The knowledge of this dictionary  direct us to a  number of avenues for further research.
These include, for instance, straightforward generalisations, such as inclusion of backbone and even sidechain atoms. The exploration of the linear
representation of folding patterns, and its potential correlation with
sequence signals is a particularly attractive challenge. Working out
the grammar associated with sequences of classes of dictionary fragments
can illuminate different levels of folding architectures. The
dictionary fragments themselves might well be applicable in approaches to predictions of protein structure, the holy grail of bioinformatics.

\bibliographystyle{IEEEtran}
\bibliography{dictms}
\end{document}